\documentclass[12pt]{iopart}
\usepackage{graphicx}

\begin{document}

\title[Universal behavior of
dispersion forces in the
low-temperature limit ]{Universal behavior of 
dispersion forces between two dielectric plates in
the low-temperature limit 
}
\author{
G~L~Klimchitskaya$^{1,2}$,
B~Geyer$^2$
and V~M~Mostepanenko$^{2,3}$}

\address{$^1$
North-West Technical University, Millionnaya St. 5,
St.Petersburg, Russia
}
\address{$^2$
Center of Theoretical Studies and Institute for Theoretical Physics,
Leipzig University, Augustusplatz 10/11, D-04109 Leipzig, Germany 
}
\address{$^3$
Noncommercial Partnership  ``Scientific Instruments'', Moscow, Russia}

\begin{abstract}
The universal analytic expressions in the limit of low
temperatures (short separations) are obtained for the free
energy, entropy and pressure between the two parallel plates
made of any dielectric. The analytical proof of the Nernst
heat theorem in the case of dispersion forces acting between
dielectrics is provided. This permitted us to formulate the
stringent thermodynamical requirement that must be satisfied
in all models used in the Casimir physics.
\end{abstract}

\section*{}
It is common knowledge that dispersion force is a quantum
phenomenon which results from fluctuating (thermal) electromagnetic
fields \cite{1}. The most well known examples of dispersion
interaction are the
van der Waals \cite{1} and the Casimir [2--4] forces.
Dispersion forces play a very important role in
surface phenomena, layered structures, colloid-substrate
interactions, adhesion, foam formation and in stability
of microelectromechanical systems [5--8].
Recently they were found to be of considerable significance
in experiments on quantum reflection and Bose-Einstein
condensation of ultracold atoms near different surfaces 
 \cite{11,12}. 
Dispersion forces create a free-energy difference between
materials in the normal and the superconducting phase which may
influence the value of the critical magnetic field \cite{14}.
They are responsible for the interaction of atoms and
molecules with nanostructures like carbon nanotubes \cite{16}.

The theoretical description of all above-listed  
phenomena is based on the Lifshitz theory \cite{18}.
This theory presents the dispersion force, 
free energy and entropy between the two
plates in terms of their dielectric permittivity
$\varepsilon(i\xi)$ along the entire
imaginary frequency axis including zero frequency. 
$\varepsilon(i\xi)$ is
found by means of the dispersion relation using the experimental
optical data for the complex refractive index \cite{20}.
As these data are available only within a
restricted frequency region, the use of some theoretical models
of dielectric response becomes unavoidable. 
Different extrapolations of data outside the
regions where they are measured (e.g., to low frequencies) may
lead, however, to very different theoretical predictions.
This places strong emphasis on the extrapolation problem in 
applications of the Lifshitz theory.

In this paper we present an analytical derivation for the
low-temperature (short-separation) behavior of the Lifshitz 
entropy and thermal corrections to the energy and
pressure between two thick dielectric plates
(semispaces). It is shown to be the same as for metals, i.e.,
universal. We demonstrate that if the plate
material  at low frequencies is described by the static 
dielectric permittivity, 
the entropy goes to zero in the limit of zero temperature
in accordance with the Nernst heat theorem. Alternatively, if
one includes the nonzero dc conductivity of the dielectric
material into the model of the dielectric
response, the entropy goes to a nonzero positive value when the
temperature $T$ goes to zero (i.e., the Nernst heat theorem is
violated). 
Finally, we formulate the thermodynamic constraint on
the extrapolations of the optical data to low frequencies 
and apply it to the topical problem of noncontact
atomic friction [15--18].

The Lifshitz formula for the free energy of
dispersion interaction between two thick plates in thermal
equilibrium, written in terms of dimensionless variables, is
\begin{equation}
\fl
{\cal F}(a,T)=\frac{\hbar c \tau}{32\pi^2a^3}
\sum\limits_{l=0}^{\infty}\left(1-\frac{\delta_{l0}}{2}\right)
\int_{\tau l}^{\infty}ydy{\!}{\!}
\left[\ln\left(1-{r_{\|}^2}{e^{-y}}\right)
+
\ln\left(1-{r_{\bot}^2}{e^{-y}}\right)\right].
\label{eq1}
\end{equation}
\noindent
Here, $a$ is the separation between the plates and we use the
dimensionless variable $\tau=4\pi k_BaT/(\hbar c)$  
($k_B$ is the Boltzmann
constant). The reflection coefficients for the two independent
polarizations of the electromagnetic field are given by 
\begin{equation}
r_{\|}=\frac{\varepsilon_ly-
\sqrt{y^2+\zeta_l^2(\varepsilon_l-1)}}{\varepsilon_ly+
\sqrt{y^2+\zeta_l^2(\varepsilon_l-1)}},
\quad
r_{\bot}=\frac{\sqrt{y^2+\zeta_l^2(\varepsilon_l-1)}-
y}{\sqrt{y^2+\zeta_l^2(\varepsilon_l-1)}+y}.
\label{eq2}
\end{equation}
\noindent
The dimensionless Matsubara frequencies are 
$\zeta_l=\xi_l/\xi_c=l\tau$ where the dimensional ones are
$\xi_l=2\pi k_BTl/\hbar$ and $\xi_c=c/(2a)$. 
The dielectric permittivity is
computed at imaginary Matsubara frequencies
$\varepsilon_l=\varepsilon(i\xi_l)=\varepsilon(i\zeta_l\xi_c)$.

Applying the Abel-Plana formula \cite{6,20}
\begin{equation}
\sum\limits_{l=0}^{\infty}\left(1-\frac{1}{2}\delta_{l0}\right)
F(l)=\int_{0}^{\infty}F(t)dt
+i\int_{0}^{\infty}dt\frac{F(it)-F(-it)}{e^{2\pi t}-1},
\label{eq3}
\end{equation}
\noindent
we can rearrange Eq.~(\ref{eq1}) to the form
$
{\cal F}(a,T)=E(a)+\Delta {\cal F}(a,T)
$
where
\begin{eqnarray}
&&
E(a)=\frac{\hbar c }{32\pi^2a^3}
\int_{0}^{\infty}d\zeta
\int_{\zeta}^{\infty}dyf(\zeta,y),
\label{eq5} \\
&&
f(\zeta,y)=y\left\{\ln\left[1-r_{\|}^2(\zeta,y)e^{-y}\right]+
\ln\left[1-r_{\bot}^2(\zeta,y)e^{-y}\right]\right\}
\nonumber
\end{eqnarray}
\noindent
is the energy of dispersion interaction at zero temperature,
and 
\begin{equation}
\Delta {\cal F}(a,T)=\frac{i\hbar c\tau}{32\pi^2a^3}
\int_{0}^{\infty}{\!\!\!}dt\frac{F(it\tau)-F(-it\tau)}{e^{2\pi t}-1},
\quad
F(x)\equiv\int_{x}^{\infty}{\!\!\!}dy f(x,y)
\label{eq6}
\end{equation}
\noindent
is the thermal correction to it.
The asymptotic expansions of the energy (\ref{eq5}) at both short
separations  and
large separations  are
well known \cite{18,20}. Here we obtain the low-temperature
(short-separation) behavior of the thermal correction (\ref{eq6})
for the case of dielectric plates.

To solve this problem, it is sufficient to describe the dielectric by its 
static dielectric permittivity $\varepsilon_0=\varepsilon(0)$.
The reason is that for dielectrics at sufficiently low temperatures
the Matsubara frequencies giving the leading contribution to Eq.~(\ref{eq6})
belong to the region where $\varepsilon$ practically does not
depend on the frequency and is equal to $\varepsilon_0$ [this is true
for $\Delta{\cal F}$ but not for $E(a)$]. To obtain the
asymptotic behavior of $\Delta {\cal F}(a,T)$ at $\tau\ll 1$
we, first, expand the function $f(x,y)$,
defined in Eq.~(\ref{eq5}), in powers of $x=t\tau$. 
Then we introduce the new variable $\tilde{y}=y-x$ to exclude $x$
from the lower integration limit in Eq.~(\ref{eq6}).
The subsequent
integration of the obtained expansion with respect to 
$\tilde{y}$ from $0$ to
infinity leads to
\begin{equation}
F(ix)-F(-ix)=i\pi\frac{(\varepsilon_0-1)^2}{2(\varepsilon_0+1)}x^2
-i\alpha x^3+O(x^4),
\label{eq8}
\end{equation}
\noindent
where $\alpha$ is real and remains unknown at this
stage because all powers of the expansion of $f(x,y)$ contribute
to its value. Next, we substitute Eq.~(\ref{eq8}) in Eq.~(\ref{eq6})
with the result
\begin{equation}
{\cal F}(a,T)=E(a)-\frac{\hbar c }{32\pi^2a^3}
\left[\frac{\zeta(3)(\varepsilon_0-1)^2}{8\pi^2(\varepsilon_0+1)}
\tau^3
\label{eq9}
-C_4\tau^4+O(\tau^5)\right],
\end{equation}
\noindent
where $C_4=\alpha/240$ and $\zeta(z)$ is the Riemann zeta-function.
Note that this equation (and respective equations for a pressure and 
entropy) does not allow a limiting transition $\varepsilon_0\to\infty$
in order to obtain the case of ideal metals. The mathematical
reason is that in our perturbation
theory it is impermissible to interchange the limits $\tau\to 0$
and $\varepsilon_0\to\infty$ in the power expansions of functions 
depending on $\varepsilon_0$ as a parameter.

The pressure of the dispersion interaction is given by
\begin{equation}
P(a,T)=-\frac{\partial{\cal F}(a,T)}{\partial a}
=P_0(a)-
\frac{\hbar c }{32\pi^2a^4}
\left[C_4\tau^4+O(\tau^5)\right],
\label{eq10}
\end{equation}
\noindent 
where $P_0=-\partial E/\partial a$ is the pressure at $T=0$ and only the
fourth-power term on the right-hand side of Eq.~(\ref{eq9})
contributes to the thermal correction. At low temperatures this
analytical result agrees with the behavior of the Casimir pressure for
nondispersive dielectrics calculated numerically in Ref.~\cite{Brev}.

Alternatively, the pressure can be found directly from the
Lifshitz formula 
\begin{equation}
P(a,T)=-\frac{\hbar c \tau}{32\pi^2a^4}
\sum\limits_{l=0}^{\infty}\left(1-\frac{1}{2}\delta_{l0}\right)
\int_{\tau l}^{\infty}y^2dy
\left[\frac{r_{\|}^2}{e^{y}-r_{\|}^2}+
\frac{r_{\bot}^2}{e^{y}-r_{\bot}^2}\right].
\label{eq11}
\end{equation}
\noindent
Applying the Abel-Plana formula (\ref{eq3}) in Eq.~(\ref{eq11}),
we get
$P(a,T)=P_0(a)+\Delta P(a,T)$
where the thermal correction to the pressure is
\begin{equation}
\Delta {P}(a,T)=-\frac{i\hbar c\tau}{32\pi^2a^4}
\int_{0}^{\infty}dt\frac{\Phi(it\tau)-\Phi(-it\tau)}{e^{2\pi t}-1}
\label{eq13}
\end{equation}
\noindent
and the function $\Phi(x)=\Phi_{\|}(x)+\Phi_{\bot}(x)$ is defined by
\begin{equation}
\Phi_{\|,\bot}(x)=\int_{x}^{\infty}
\frac{y^2dy\,r_{\|,\bot}^2(y,x)}{e^{y}-r_{\|,\bot}^2(y,x)}\,.
\label{eq14}
\end{equation}

By finding the leading term of the expansion of $\Phi(x)$ 
in powers of $x$, one arrives at
\begin{equation}
\Phi(ix)-\Phi(-ix)=-i\frac{x^3}{3}(\sqrt{\varepsilon_0}-1)
(\varepsilon_0^2+\varepsilon_0\sqrt{\varepsilon_0}-2)+O(x^5).
\label{eq18}
\end{equation}

Substitution of Eq.~(\ref{eq18}) into Eq.~(\ref{eq13}) leads to the
result
\begin{equation}
P(a,T)=P_0(a)-\frac{\hbar c}{32\pi^2a^4}
\left[\frac{(\sqrt{\varepsilon_0}-1)
(\varepsilon_0^2+\varepsilon_0\sqrt{\varepsilon_0}-2)}{720}
\tau^4+O(\tau^5)\right].
\label{eq19}
\end{equation}
\noindent
Comparing Eqs.~(\ref{eq10}) and (\ref{eq19}) we find
the value of so long unknown coefficient
\begin{equation}
C_4=\frac{1}{720}(\sqrt{\varepsilon_0}-1)
(\varepsilon_0^2+\varepsilon_0\sqrt{\varepsilon_0}-2).
\label{eq20}
\end{equation}
\noindent
Thus, the low-temperature (short-separation) behavior of both the free
energy and the pressure is given by  Eqs.~(\ref{eq9}), (\ref{eq19}),
(\ref{eq20}). 
By using these results, the asymptotic behavior 
of the entropy of dispersion interaction is described by the expression
\begin{eqnarray}
&&
S(a,T)=-\frac{\partial{\cal F}(a,T)}{\partial T}=
\frac{3k_B\zeta(3)(\varepsilon_0-1)^2}{64\pi^3a^2(\varepsilon_0+1)}
\tau^2
\label{eq21} \\
&&\phantom{aaa}
\times\left[1-
\frac{2\pi^2(\varepsilon_0+1)(\varepsilon_0\sqrt{\varepsilon_0}+
2\varepsilon_0+2\sqrt{\varepsilon_0}+
2)}{135\zeta(3)(\sqrt{\varepsilon_0}+1)^2}\tau\right].
\nonumber
\end{eqnarray}
\noindent
We see from  Eq.~(\ref{eq21}) that in the limit $\tau\to 0$
($T\to 0$) the entropy of both the van der Waals and Casimir
interactions goes to zero following the same universal law which
was previously found for ideal  and for real
metals \cite{20}.
We have proved that the use of the Ninham-Parsegian representation
\cite{1} for $\varepsilon(i\xi_l)$ instead of $\varepsilon_0$
modifies only the terms of order $O(\tau^5)$ in
Eqs.~(\ref{eq9}), (\ref{eq19}).  
The comparison with the results of numerical computations for
real dielectrics demonstrates that at separations
100--500\,nm our asymptotic expressions are applicable at
$T<60-70\,$K.

We now turn to a problem of major importance which arises when 
one includes  the dc conductivity of the dielectric plates
into the model of the dielectric response,
$
{\tilde{\varepsilon}}_l={\varepsilon}_l+{4\pi\sigma_0}/{\xi_l}
={\varepsilon}_l+{\beta(T)}/{l}.
$
Here $\sigma_0$ is the dc conductivity of the dielectric and
$\beta=2\hbar\sigma_0/(k_BT)$. The conductivity depends on 
$T$ according to $\sigma_0\sim\exp(-b/T)$ where $b$
is different for different
dielectrics. It is significant that for
dielectrics the additional Drude term 
 is very small for all $\xi_l\neq 0$.
For example, 
$\beta\sim 10^{-12}$ for
SiO${}_2$ at $T=300\,$K and, thus, it is for sure negligible
for all $l\geq 1$.

One might believe, however, that this term plays a role in the
zero-frequency contribution in Eq.~(\ref{eq1}). To test
this conjecture we substitute ${\tilde{\varepsilon}}_l$ 
in Eq.~(\ref{eq1}) and arrive at
\begin{equation}
{\tilde{\cal F}}(a,T)={\cal F}(a,T)-
\frac{k_BT}{16\pi a^2}\left\{\zeta(3)-\mbox{Li}_3
\left[\left(\frac{\varepsilon_0-1}{\varepsilon_0+1}\right)^2
\right]
\vphantom{\left[\left(\frac{\varepsilon_0}{\varepsilon_0}\right)^2
\right]
}+R(\tau)\right\},
\label{eq23}
\end{equation}
\noindent
where Li${}_3(z)$ is the polylogarithm function, the asymptotic
behavior of ${\cal F}$ is given by Eqs.~(\ref{eq9}), (\ref{eq20}),
and $R$ decreases exponentionally when $T\to 0$. As a result the
entropy of the dispersion interaction at $T=0$,
\begin{equation}
{\tilde{S}}(a,0)=
\frac{k_B}{16\pi a^2}\left\{\zeta(3)-\mbox{Li}_3
\left[\left(\frac{\varepsilon_0-1}{\varepsilon_0+1}\right)^2
\right]\right\}>0,
\label{eq24}
\end{equation}
\noindent
in violation of the Nernst heat theorem. Thus, the dc conductivity
of a dielectric must not be included in the models
of dielectric response. This should be compared
with the case of plates made of real metal
(see Refs.~\cite{Brev,32} and review \cite{35} for details),
where different opinions on the validity of the Nernst heat theorem
were proposed. In fact, the mechanisms for the violation of this
theorem in some models of metals and dielectrics are quite different. 
In metals, the validity of the Nernst heat theorem depends on the
scattering processes of free charge carriers on phonons, defects
(impurities) etc. For the Drude metals with impurities (like in
Ref.~\cite{Brev}) the residual relaxation at $T=0$ is not equal
to zero and the Nernst heat theorem is satisfied. The same takes place
in the case of metals described by the plasma model. For perfect
crystal lattices of the Drude metals with no impurities, relaxation at
zero temperature is absent and the Nernst heat theorem is violated
\cite{32}. All these cases are discussed in Ref.~\cite{status} devoted
to metals. For dielectrics, the validity of the Nernst heat theorem
does not depend on the scattering processes due to quick vanishing 
of the concentration of carriers when the temperature vanishes. Here,
the violation occurs due to the inclusion of infinitely large dielectric
permittivity at zero frequency. Even the sign of the entropy at zero
temperature for metals and dielectrics is opposite
(negative for perfect crystal lattices of the Drude metals and
positive for dielectrics with included dc conductivity).
For a complete discussion of this subject, containing all mathematical
details, see Ref.~\cite{PRD05} where Eq.~(\ref{eq24}) is re-derived in
the framework of a more general case of two dissimilar dielectrics.

The above results are important for many
applications of dispersion forces. As an example
we refer to the problem of a noncontact atomic friction where
the discrepancy between experiment and theory is very large 
\cite{24,27}. In Ref.~\cite{27} it has been proposed
that the friction observed in the experiment of Ref.~\cite{24}
could be due to the dc conductivity of an underlying SiO${}_2$  
plate described by ${\tilde{\varepsilon}}_l$.
From the preceding discussion, it can be seen that
such a proposition would not be in agreement with 
the thermodynamic constraint. Further applications of this 
constraint in the theory of dispersion forces are under way
(see Ref.~\cite{PRD05} related to the case of dissimilar
dielectrics). 

\section*{Acknowledgments}
G.L.K. is grateful to L.~P.~Pitaevskii for attracting her
attention to this problem.
This work was supported by Deutsche Forschungsgemeinschaft grant 
436\,RUS\,113/789/0-1.


\end{document}